\documentclass[preprint, review, 12pt]{elsarticle}

\usepackage{amssymb}
\usepackage{amsmath}
\usepackage{lineno}
\usepackage{ulem}
\usepackage{natbib}
\usepackage{hyperref} 
\usepackage{lineno}
\usepackage{makecell}
\usepackage{subfig}
\journal{Journal of }

\begin{document}

\begin{frontmatter}

%% Title, authors and addresses

%% use the tnoteref command within \title for footnotes;
%% use the tnotetext command for theassociated footnote;
%% use the fnref command within \author or \affiliation for footnotes;
%% use the fntext command for theassociated footnote;
%% use the corref command within \author for corresponding author footnotes;
%% use the cortext command for theassociated footnote;
%% use the ead command for the email address,
%% and the form \ead[url] for the home page:
%% \title{Title\tnoteref{label1}}
%% \tnotetext[label1]{}
%% \author{Name\corref{cor1}\fnref{label2}}
%% \ead{email address}
%% \ead[url]{home page}
%% \fntext[label2]{}
%% \cortext[cor1]{}
%% \affiliation{organization={},
%%             addressline={},
%%             city={},
%%             postcode={},
%%             state={},
%%             country={}}
%% \fntext[label3]{}

\title{Linking Calendar and Cycle Ageing in Lithium-Ion Batteries through Consistent Parameterisation of an Electrochemical–Thermal–Degradation Model} %% Article title

%% use optional labels to link authors explicitly to addresses:
%% \author[label1,label2]{}
%% \affiliation[label1]{organization={},
%%             addressline={},
%%             city={},
%%             postcode={},
%%             state={},
%%             country={}}
%%
%% \affiliation[label2]{organization={},
%%             addressline={},
%%             city={},
%%             postcode={},
%%             state={},
%%             country={}}

\author{Ganesh Madabattula\corref{cor1}} %% Author name
\cortext[cor1]{Corresponding author: ganesh.che@iitbhu.ac.in}
%% Author affiliation
\affiliation{organization={Department of Chemical Engineering and Technology, Indian Institute of Technology (BHU) Varanasi},%Department and Organization
            %addressline={}, 
            city={Varanasi},
            postcode={221005}, 
            state={Uttar Pradesh},
            country={India}}

%% Abstract
\begin{abstract}
%% Text of abstract briefly state the purpose of your research, principal results and major conclusions.
  Parameterisation of coupled degradation mechanisms in lithium-ion batteries is a major challenge. Interactions between the mechanisms depend on usage conditions: C-rate, rest state-of-charge (SoC), depth-of-discharge (DoD) and temperature. This work presents a framework to consistently parameterise key degradation modes—solid-electrolyte interphase (SEI) growth, lithium plating, and active material loss in both electrodes—using insights derived from degradation mode analysis data. The work predicts capacity fade trajectories of a  NMC-based lithium-ion cell under both calendar and combined calendar-cyclic ageing, using a P2D electrochemical–thermal–degradation model. The work predicts state-of-health (SoH), remaining-useful-life (RUL) and internal degradation modes of the cell--under 81 combinations of  temperature (10$^o$C, 25$^o$C, 40$^o$C), C-rate (0.1 C, 0.3 C and 1.0 C), rest SoC (10\%, 60\%, and 100\%) and DoD  (50\%, 70\%, and 90\%)--using PyBaMM. The predicted cycle-life varies between 0.8 to 14 years  to reach 75\% of SoH.  The work provides mechanistic insights into competing effects between calendar and cyclic ageing, during cycling.  The model demonstrates sub-linear, linear, and sup-linear/accelerated capacity fade based on the usage conditions. The simulated dataset for all the cases is made available.

\end{abstract}

%%Graphical abstract
% \begin{graphicalabstract}
% %\includegraphics{grabs}
% \end{graphicalabstract}

%%Research highlights
% \begin{highlights}

% \item Predicted ageing trajectories under different modes using a electrochemical-thermal-degradation model.
% \item SoC, temperature, DoD and C-rate dependence on cyclic/calendar ageing captured.
% \item Four (SEI, plating, loss active material) coupled degradation mechanisms are consistently parameterised using a degradation mode analysis insight.
% \item SoH, RUL and internal degradation modes for 81 usage cases simulated using PyBaMM.
% \item Quantified competing effects between cyclic and calendar ageing during cycling with rest.

% \end{highlights}

%% Keywords
\begin{keyword}
%% keywords here, in the form: keyword \sep keyword

%% PACS codes here, in the form: \PACS code \sep code

%% MSC codes here, in the form: \MSC code \sep code
%% or \MSC[2008] code \sep code (2000 is the default)

Lithium-ion batteries \sep consistent degradation trajectories \sep electrochemical-thermal-degradation modelling \sep accelerated capacity fade \sep calendar ageing \sep cyclic ageing \sep state-of-health

\end{keyword}

\end{frontmatter}

%% Add \usepackage{lineno} before \begin{document} and uncomment 
%% following line to enable line numbers
%% \linenumbers

%% main text
%%

%% Use \section commands to start a section
\section{Introduction}
\label{sec1}
%% Labels are used to cross-reference an item using \ref command.
Imagine a bunch of bananas presented to a group of hungry monkeys. How many bananas remain after an hour? How many does each monkey eat? The answers defy simple prediction, hinging on factors like mood, hunger levels, and social dynamics.
Now consider a lithium-ion cell with initial capacity Q$_0$ Ah, where degradation mechanisms—our "monkeys"—voraciously consume capacity over time. How much capacity survives after a year of use? What share does each degradation mechanism claim? How many years can the cell endure under varying operating conditions? This paper addresses these questions through predictions from a physics-based model. 

\begin{figure}
    \centering
    \includegraphics[width=0.8\linewidth]{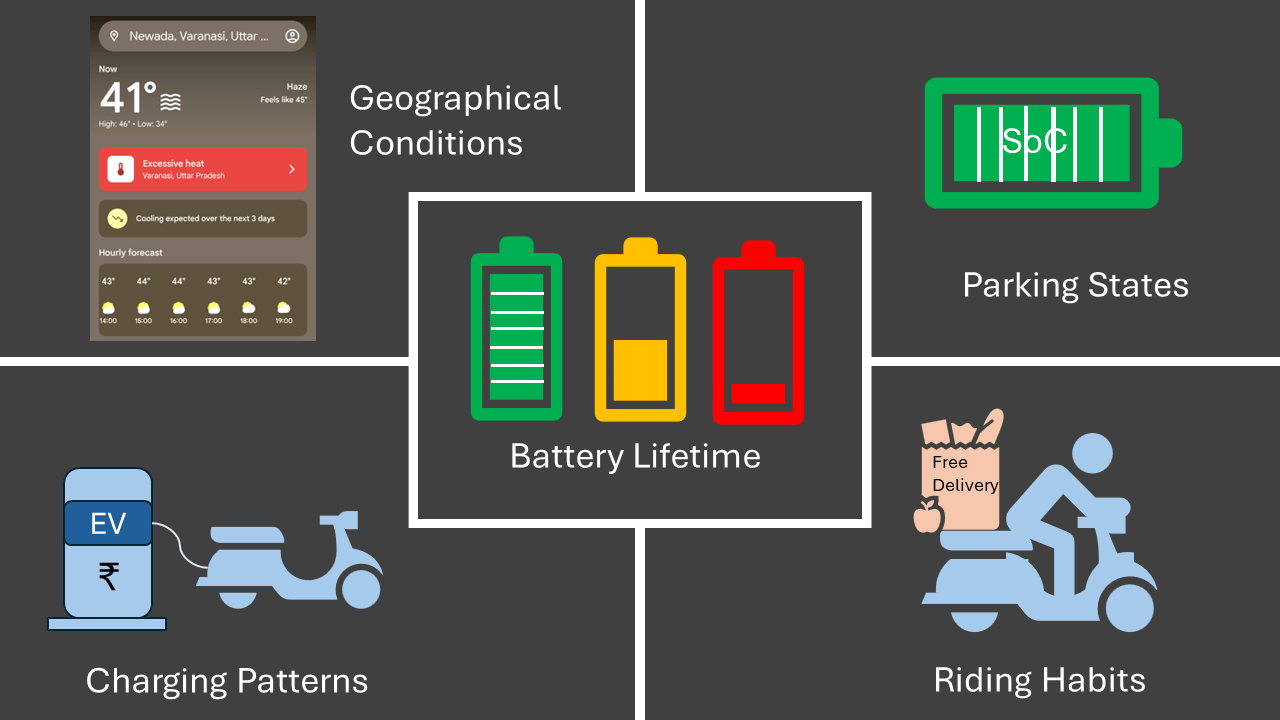}
    \caption{A schematic representation of usage conditions that effect battery lifetime: temperature, rest SoC, DoD and C-rate.}
    \label{fig:schematicDod}
\end{figure}

Lithium-ion batteries (LIBs) are widely used in e-mobility and energy storage due to their high energy density and pragmatic cycle life. Cycle-life of the battery depends on usage patterns and conditions such as C-rate, state-of-charge (SoC) at rest, depth-of-discharge (DoD) and ambient temperature, as summarised in Figure~\ref{fig:schematicDod} \cite{Birkl2017,rahman2024exploring,Edge2021}.  Having an accurate estimate of available capacity and remaining useful life of a cell improves the application’s reliance, consumer experience, battery safety, and cycle-life. 

Prediction of  remaining useful life (RUL) and state-of-health (SoH) of the cell is challenging due to difficulty in tracking complex interactions of degradation causing mechanisms inside the cell.\cite{Birkl2017,OKane2022} The coupled interactions vary with usage conditions: SoC, DoD, C-rate, temperature, chemistry and mode of operation. Moreover, the cells vary in size, shape and their chemistry, so predictions on one cell cannot be linearly extrapolated to another cell. Solid-electrolyte interface (SEI) reaction \cite{aurbach1995study, aurbach2009surface, verma2010review, wang2016direct,Kamyab2019MixedSEI, Kamyab2019}, lithium-plating at anode \cite{koleti2020new, uhlmann2015situ, arora1999mathematical,liu2016understanding,waldmann2018li,ge2017investigating,legrand2014physical}, transition metal dissolution at cathode \cite{allen2024coordination, zhan2018dissolution, hellar2025direct, Zhan2018DissolutionReview,gajan2025dynamics} and loss of active material due to mechanical fracture \cite{li2017situ, warburton2020oriented, christensen2006stress,zhang2007numerical,reniers2019review,ai2020electrochemical, laresgoiti2015modeling, li2018single} at both the electrodes during cycling are the major degradation mechanisms in the batteries \cite{vetter2005ageing, hendricks2015failure,han2019review, Edge2021,jeon2025state}. \citet{Birkl2017} and \citet{Edge2021} summarised the degradation mechanisms, the coupled effects between these mechanisms, and the influence of usage conditions on the mechanisms in detail. The works also provide excellent schematics of the mechanisms and their cause-and-effect interactions. 

A comprehensive understanding on the effect of various combinations of usage conditions on capacity fade, which internally affect rates of the coupled degradation mechanisms and internal states is quite useful. Such study can help  extend its cycle life and help have accurate estimate of SoH on-board. Conducting such experimental study is impractical to account for variation in design parameters, usage modes and conditions and in-situ tracking of degradation mechanism w.r.t time and investment. Fortunately, this knowledge gap can be filled by a physics-based modelling study. An review of physics-based modelling of degradation in the cells is carried out and summarised in \citep{reniers2019review}.   However, a very few works attempted a comprehensive physics-based model study focussed on degradation which can elucidate the effect of different combinations of usage conditions on cycle-life under different ageing modes, with multiple degradation mechanisms accounted, in the literature, to our knowledge. 

Previous works\cite{ OKane2022,reniers2019review,li2025importance}, which coupled three or more degradation mechanisms in the cell and modelled them form the basis for the current work to provide insights on capacity fade. \citet{reniers2019review} provide the comprehensive summary of model equations for different degradation mechanisms and presents simulation trends of degradation under cyclic ageing using a single particle model (SPM). \citet{OKane2022} modelled coupled degradation mechanisms in the cell using PyBaMM \cite{Sulzer2020} and highlighted the importance of parameterisation for consistent degradation modelling . By extending the physics based P2D model \cite{Doyle1993ModelingCell}, the authors carried out parametric study on the trends of degradation under cyclic ageing and the coupled effect between the mechanisms for the first time. Their model is isothermal. Recent work by \citet{li2025importance} attempts to model the degradation in NMC cell at different temperatures for specified DoD limits and highlights the use of degradation mode analysis for parameterisation of the mechanisms. The authors coupled 5 degradation mechanisms in a P2D model.

The current work focuses on parameterising the model consistently by combining  model equations for SEI layer growth, lithium plating and active material loss in both electrodes and predict cycle-life at different temperatures, C-rates, DoD and rest SoC, using a P2D model and discuss parameterisation strategies. Then, we demonstrate degradation trajectories for calendar, cycling and combined ageing under different combinations of usage conditions using PyBaMM code, which has not been attempted yet in the literature.\cite{Sulzer2020} We look into the capacity loss contributions of each mechanisms in the ageing cases using simulated degradation mode analysis profiles. We predict the capacity fade in calendar ageing mode at 9 combinations of temperatures and rest SoCs.  Under combined cyclic and calendar ageing, we predict capacity fade trajectories under 81 combinations of usage variables: temperature (10$^o$C, 25$^o$C, 40$^o$C), rest SoC (10\%, 60\%, and 100\%), DoD (50\%, 70\%, and 90\%), and C-rate (0.1 C, 0.3 C and 1.0 C) together, and their effect on coupled interactions on the degradation mechanisms. We discuss the competing effects between calendar and cycling ageing during the combined mode. 

A modelling work that demonstrates calendar, cyclic and combined ageing trends, under different combinations of usage conditions, using a P2D physics-based model and provides insights on degradation modes is not yet available in the literature, to our knowledge. The work provides insights on coupled effects of usage conditions on the coupled degradation mechanisms to predict SoH and RUL. These insights can help validate the battery models under different modes of operation for consistency check and improve cell design for extended cycle life. The analysis also helps make informed system integration of the cells in applications for better battery management.

\section{Details of the modelling}
%% Inline mathematics is tagged between $ symbols.
The electrochemical-thermal-degradation model for the lithium-ion cell is based on the Doyle-Fuller-Newman (DFN) model \cite{Doyle1993ModelingCell, Chen2020} with additional degradation mechanisms and lumped thermal model added to it\cite{OKane2022, reniers2019review}. The Python Battery Mathematical Modelling (PyBaMM), an open-source code \cite{Sulzer2020}, for the DFN model with the relevant degradation mechanisms is used for the simulations in this work. The DFN model equations without degradation can be found on the PyBaMM documentation url \cite{pybamm_dfn_docs}. The parameters for the LGM50 cell with NMC811 cathode and composite anode (Graphite/Si) are used for the model. The parameters set, \textit{OKane2020} \cite{OKane2022, pybamm_OKane2022_params}, for the cell, available on PyBaMM, is used for the simulations, which has degradation parameters. The parameters are originally based on \textit{Chen2020} parameter set \cite{Chen2020}, created for beginning of life performance modelling of LG M50 cell. \textit{Chen2020} parameters set without ageing has been validated earlier with the experiments by \citet{Chen2020}. Later, \citet{OKane2022} included degradation parameters and developed a coupled modelling framework on PyBaMM to simulate qualitative capacity fade trends.  Following on the earlier works \cite{OKane2022, Chen2020, ORegan2022}, we also assume the negative electrode as a composite and treat it as a single active material with composite properties, to focus on demonstrating consistent capacity fade trends under different usage modes. For the current work, the assumption holds reasonable to demonstrate consistency of capacity fade trends in different modes qualitatively.

For the degradation, we include solid-electrolyte interphase (SEI) reaction at anode, irreversible lithium plating at anode and active material loss due to mechanical stresses at both electrodes. The equations for the mechanisms are presented as follows.

\subsection{Equations for capacity fade mechanisms}
\subsubsection{Solid-electrolyte interphase loss}
The capacity loss due to the reaction between charged negative electrode and solvent to form SEI is correlated with the SEI current ($i_{sei}$)  as per the equation \cite{Kamyab2019MixedSEI, reniers2019review,safari2009multimodal}:

\begin{equation}
i_{sei} = \frac{c_{s,bulk}}{\frac{1}{nFk_{sei}exp(-\frac{\alpha nF}{RT}\eta_{sei})} + \frac{\delta}{nF D_{sei}}}.
\end{equation}
Here, $c_{s,bulk}$ is the solvent concentration, $k_{sei}$ is the rate constant for SEI reaction with $\eta_{sei}$ as the over-potential across the interphase. $D_{sei}$ is the diffusivity of the solvent through the SEI layer and $n$ is the number of electrons involved in SEI reaction.
The change in SEI layer thickness ($\delta$) is given as
\begin{equation}
\frac{\partial \delta}{\partial t} = \frac{i_{sei}M_{sei}}{\rho_{sei} nF}
\end{equation}
where $M_{sei}$ is the molecular weight and $\rho_{sei}$ is the density of the SEI layer. The mechanism is considered to be both diffusion and reaction limited (mixed). A pure diffusion limited mechanism would not be sufficient to account for potential dependence on calendar ageing. 

\subsubsection{Lithium plating}
Lithium plating causes loss of lithium inventory and safety issues. For irreversible plating, following the Butler-Volmer kinetics \cite{newman2021electrochemical, bard2022electrochemical} for an irreversible reaction, the plating current\cite{OKane2022,reniers2019review, okane2020physical,wood2016dendrites} is given as:
\begin{equation}
i_{pl} = i_{0,pl}exp\left( \frac{\alpha nF}{RT}\eta_{pl}\right),
\end{equation}
% \begin{equation}
% \eta_{pl} = V_{neg} + \eta_{neg} - V_{pl} + r_{sei}\delta I.
% \end{equation}
where $i_{0,pl}$ is the exchange current density, dependent on electrolyte concentration and $\eta_{pl}$ is over-potential for for plating reaction. The overpotential includes potential drop across the SEI layer and links it with the growing SEI layer. Li plating dominates at a combination of colder temperatures and high C-rates, where lithium diffusion dynamics are slowed down and metal reduction occurs. For current study, we consider irreversible plating only and ignore stripping. 
\subsubsection{Loss of active material}
The stresses due to concentration gradients in the particles during cycling cause loss in active material fraction and thereby capacity fade. The loss of active material (LAM) due to mechanical stresses is given by \cite{OKane2022,zhang2007numerical, laresgoiti2015modeling, dai2014simulation,bohn2013model, li2017single,fu2013modeling}:
\begin{equation}
    \frac{\partial \varepsilon_a}{\partial t} = \frac{\beta}{t_0}\left( \frac{\sigma_h}{\sigma_c}\right)^{m_2}        \qquad \sigma_h > 0,           
\end{equation}
where $\beta$ and $m_2$ are two fitting constants from the experiments. $\varepsilon_a$ is active material volume fraction; The change in volume fraction affects the interfacial current density for the normal charge/discharge and couples with the SEI and the plating. $\sigma_h$ is the hydrostatic stress and $\sigma_c$ is the critical stress that bears the fracture.

\begin{equation}
    \sigma_r = \frac{2 \Omega E}{(1 - \nu)}\left[ c_{avg}(R_i) - c_{avg}(r) \right]
\end{equation}

\begin{equation}
    \sigma_t = \frac{ \Omega E}{(1 - \nu)}\left[2 c_{avg}(R_i) + c_{avg}(r) - c^*/3 \right]
\end{equation}

where \[ \sigma_h = (\sigma_r + 2\sigma_t)/3 \], $\Omega$ is partial molar volume, $E$ is Young's modulus, $\nu$ is Possion's ratio, $\sigma_r$ is radial stress, $\sigma_t$ is tangential stress, $R_i$ is radius of the particle. $c_{avg}(r)$ is the average concentration of lithium ions from centre to the radius $r$ in the particle. $c^*$ = $c$ - $c_{ref}$, $c_{ref}$ is the reference concentration for stress free case.  

We consider active material loss due to stresses in both positive and negative electrodes with different magnitudes and respective active material properties.

\subsubsection{Lumped thermal model}
Thermal behaviour of the cell is accounted through the lumped thermal equation \cite{reniers2019review,pals1995thermal, guo2011single}:
\begin{equation}
m C_p \frac{\partial T}{\partial t}= I^2R_{batt} + I(\eta_{neg}-\eta_{pos})  +  IT \frac{\partial U_{ocv}}{\partial T}  - hA_{batt}(T - T_{amb}).
\end{equation}
where we ignore the effects of thermal gradients inside a cell to keep the model practical for solving and good enough  for the current study. The first term on RHS represents heat generated to through cell resistance, second term due to charge/discharge mechanisms, and third term due to entropic losses. Fourth term represents heat loss to ambience via convective cooling with heat-transfer coefficient, $h$. $A_{batt}$ is the area for heat loss, $C_p$ is the heat capacity of the cell, $m$ is mass of the cell, $I$ is cell current, and $R_{batt}$ is the cell resistance. $\eta_{neg}$ and $\eta_{pos}$ are over-potentials at the respective electrodes. 

The above equations are combined with the DFN model for beginning of life\cite{Sulzer2020,Doyle1993ModelingCell, Chen2020, pybamm_dfn_docs} and simulated for degradation analysis.

\section{Details of the simulations}
\subsection{Beginning-of-life checks}
For all the cycling simulations, the cell is cycled between 4.2 to 2.5 V. The new cell has a discharge capacity of 5 Ah.  For capacity check, the simulated profiles at the beginning-of-life , voltage vs. discharge capacity and voltage vs. time, are shown in Figure~\ref{fig1apxBOL}, which indicate 5 Ah at 1~C. The discharge performance of the cell at different C-rates is also presented in the figure.

\subsection{Parameterisation of the model for ageing simulations}

\begin{figure}
    \centering
    \includegraphics[width=0.6\linewidth]{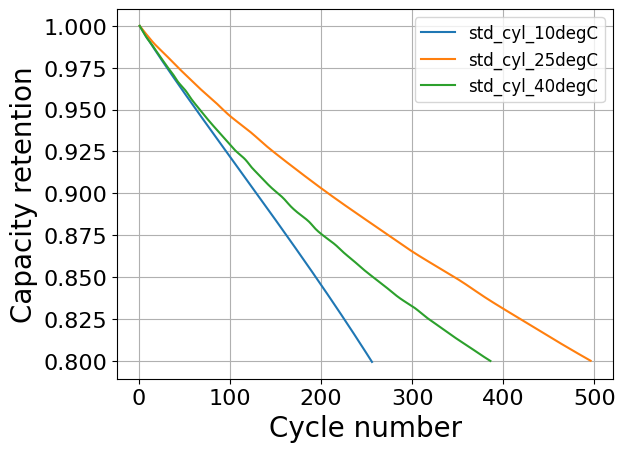}
    \caption{Prediction of capacity fade vs. cycle number during standard cycling mode at 10$^o$C, 25$^o$C and 40$^o$C. The model was parameterised to give 500 cycles for 80\% SoH at 25$^o$C, as per the data-sheet of LG M50 cell. Further, temperature dependence tuning was done to reflect the degradation mode analysis data reported by \citet{kirkaldy2024lithium} for the same cell.}
    \label{fig:102540stdcycle}
\end{figure}

For modelling degradation, the major challenge is to identify a correct set of degradation parameters that work together perfectly to every feature of capacity fade with temperature, SoC, DoD and C-rate, and usage mode.  For our study, the degradation parameters of SEI, LAM and Li plating are tuned to give 500 cycles for drop in SoH to 80\% under standard cycling conditions ( 0.33 C CC charge at 4.2 V with CV hold until 50~mA, rest for 10 min, CC discharge at 0.33 C until 2.85 V, and rest for 20 min, at 25$^o$C), according to the LG M50 data sheet (INR21700 M50T 18.2Wh, Doc. no: LRB-PS-CY18.2Wh-M50)). The predicted cycle-life under the standard cycling conditions, demonstrating 500 cycles for 80\% SoH at 25$^o$C is shown in Figure~\ref{fig:102540stdcycle}. 

Further, the figure also shows predicted capacity fade profiles at 10$^o$C and 40$^o$C under standard test conditions. The cycle-life is shortest at 10$^o$C, shorter at 40$^o$C and longest at 25$^o$C. The cycle-life is close to 386 at 40$^o$C and 256 at 10$^o$C for 80\% SoH. Our objectives for parameterisation w.r.t temperature are motivated by the degradation mode analysis of experimental data for the LG M50 cell, reported by \citet{kirkaldy2024lithium}. The parameters were tuned such that, at high temperatures (40$^o$C), capacity fade needs to be dominated by mixed-regime SEI; at normal temperatures (25$^o$C), it has to be controlled by both mixed-regime SEI and mild loss of active material; and at low temperatures (10$^o$C), the capacity loss has to be dominated by lithium plating (LLI) and higher loss of active material. At higher temperatures, SEI kinetics gets enhanced. At low temperatures, Li diffusion in active material slows down and increases over-potential for plating; the slower diffusion also leads to higher concentration gradients within the particle which generates higher stress and thereby accelerated LAM.

Fig.~7 (right columns) in \citet{kirkaldy2024lithium} shows the capacity fade data vs. charge throughput (kAh) of the cell at 10$^o$C, 25$^o$C and 40$^o$C during CC-CV charge-CC discharge cycling (referred as experiment 5) and  WLTP drivecycle discharge (referred as experiment 4),  between 0-100\% SoC. The authors observed the following: the capacity fade is fastest at 10$^o$C, faster at 40$^o$C and slowest at 25$^o$C. Their degradation mode analysis (DMA) shows faster capacity fade at 10$^o$C is contributed by higher LAM in negative electrode and higher LLI (attributed to lithium plating). LAM in negative electrode is higher than positive electrode at the three temperatures. Resistance increase is higher at 40$^o$C compared to 25$^o$C, indicating higher SEI layer thickness and the corresponding loss. LLI is higher at 40$^o$C compared to 25$^o$C, indicating higher SEI loss.  These insights formed the basis for the parameterisation of the coupled degradation mechanisms in our model. A similar approach for parameterising the model using DMA from the experimental data for ageing simulations of cycling between 70-85\% SoC and 85-100\% SoC, with a different combination of degradation mechanisms, is reported  in \citet{li2025importance}. Their parameters need to be further tuned to work for the 0-100\% SoC cycling data.

\begin{figure}
    \centering
    \includegraphics[width=1\linewidth]{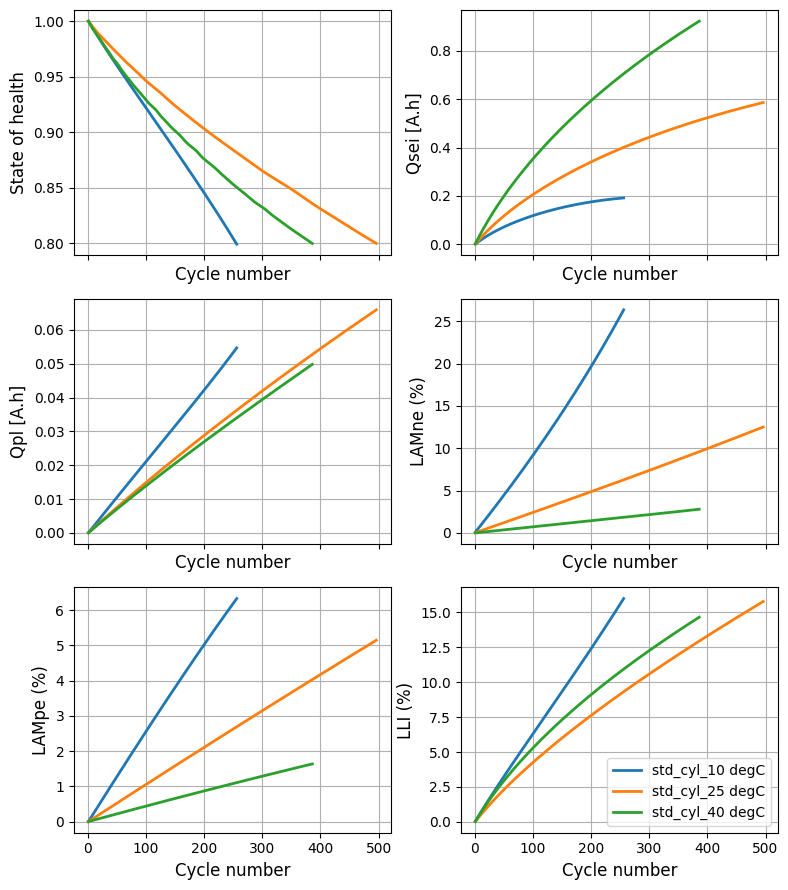}
    \caption{Simulated degradation mode analysis of the capacity fade trajectories at 10$^o$C, 25$^o$C and 40$^o$C. Capacity fade is dominated by SEI loss at 40$^o$C, and LAM and Li plating at 10$^o$C. Qsei: Capacity loss due to SEI (Ah), Qpl: capacity loss due to Li plating (Ah), LAMne: LAM in negative electrode (\%), LAMpe: LAM in positive electrode (\%), and  LLI: loss of lithium inventory (\%).}
    \label{fig:102540stdcycleDMA}
\end{figure}

Figure~\ref{fig:102540stdcycleDMA} presents the simulated DMA, using our model, corresponding to the capacity fade trajectories for 10$^o$C, 25$^o$C and 40$^o$C, under the standard cycling conditions, reported in Figure~\ref{fig:102540stdcycle}. The figure demonstrates the following: loss of capacity due to SEI ($Qsei [A.h]$) is highest at 40$^o$C and lowest at 10$^o$C; loss due to Li plating ($Qpl [A.h]$) is highest at 10$^o$C for given cycle number; LAM in negative electrode ($LAMne (\%)$) is highest at 10$^o$C and lowest at 40$^o$C; and LAM in the negative electrode is higher compared to the positive electrode ($LAMpe (\%)$) at all the temperatures. The figure also shows loss of lithium inventory ($LLI (\%)$) due to combined ageing by SEI and the plating. Lithium plating gets enhanced by the increased contribution of the potential drop across the growing SEI layer thickness to the plating over-potential.  These observations corroborate with experimental DMA results reported in \citet{kirkaldy2024lithium}, and confirm implementations of the parameterisation objectives we followed. The simulated data of capacity fade and DMA are provided in the Data Availability section. 

The degradation parameters, which were tuned for establishing consistency as mentioned above, in the current work, different from $OKane2022$ set, are listed in Table~\ref{tab:paramDegrade}.  The model options chosen on PyBaMM for the capacity fade simulations are listed in Table~\ref{tab:modeloptions}. %\uline{The code notebook of the inputs used in PyBaMM for running the simulations is provided in the supplementary section.}

\begin{table}[h!]
    \centering
    \caption{The parameters used in the simulations for the degradation mechanisms, which are different from $OKane2022$\cite{pybamm_OKane2022_params} parameter set.}
    \begin{tabular}{|l|l|}
        \hline
     Parameter     & Value  \\
     \hline
      EC diffusivity [$m^2.s^{-1}$]   & $ 6 E-20 exp\left( \frac{E_D}{R}(\frac{1}{298.15}-\frac{1}{T}) \right) $ \\
      E$_D$ [J/mol] & 10000\\
      SEI kinetic rate constant [m.s$^{-1}$] & 1E-15\\
      SEI growth activation energy [J.mol-1] & 48000.0\\
      SEI partial molar volume [m3.mol-1]& 3.834E-05\\
      Positive electrode LAM constant proportional term [s$^{-1}$] & 0.075/3600\\
      Negative electrode LAM constant proportional term [s$^{-1}$] & 0.1425/3600,\\
      Positive electrode LAM constant exponential term & 2.2\\
      Negative electrode LAM constant exponential term& 2.2\\
      Lithium plating kinetic rate constant [m.s$^{-1}$]& 1E-12\\ 
      Negative electrode OCP entropic change [V.K$^{-1}$]&-0.0002\\
      Positive electrode OCP entropic change [V.K$^{-1}$]&-0.0004\\
      Contact resistance [Ohm]&0.001\\
      Total heat transfer coefficient [W.m$^{-2}$.K$^{-1}$]& 10.0\\
      \hline
    \end{tabular}
    
    \label{tab:paramDegrade}
\end{table}

\begin{table}[h!]
    \centering
       \caption{Model options on PyBaMM for the degradation simulations.}
    \begin{tabular}{|p{4cm}|p{6cm}|}
      \hline
      Parameter set   & "OKane2022" \\
      \hline
      Model & DFN\\
\hline
Solver & IDAKLUSolver\\
\hline
      Model options   & "thermal": "lumped",

      "lithium plating": "irreversible", 

"SEI":"ec reaction limited", 

"SEI film resistance":"average", 

"SEI porosity change": "true",

"loss of active material": "stress-driven",

"particle mechanics":"swelling only"\\
\hline

    \end{tabular}
 
    \label{tab:modeloptions}
\end{table}

\subsubsection{Details of the machine used:}
A Windows-operated desktop with i9 processor, 8 GB GPU, and 128 GB RAM is used for the simulations. WSL:Ubuntu 24.04 Linux subsystem app and VS Code IDE are used to run PyBaMM (version 25.8.0) on the desktop. Python 3.12.3 is used for running the code.

\section{Results and discussion}
\label{sec:rnd}
\subsection{Calendar ageing: temperature and SoC dependence}

\begin{figure}
    \centering
    
    \subfloat{a)}{\includegraphics[width=0.6\linewidth]{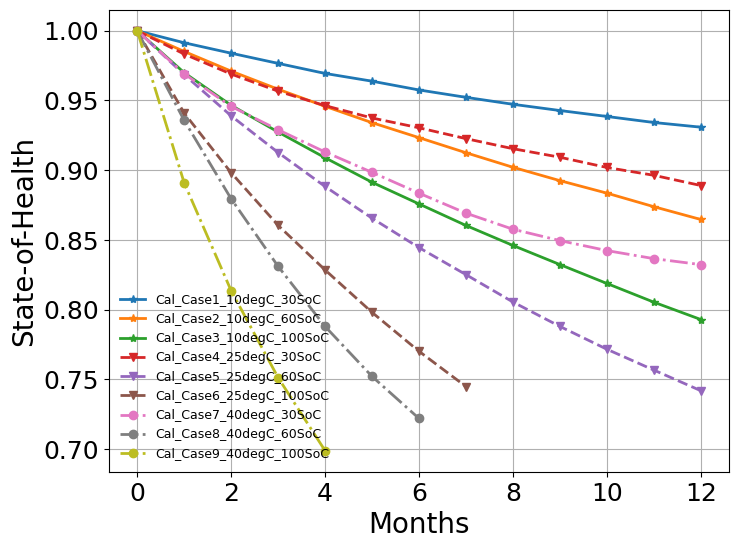}}
    \subfloat{b)}{\includegraphics[width=0.6\linewidth]{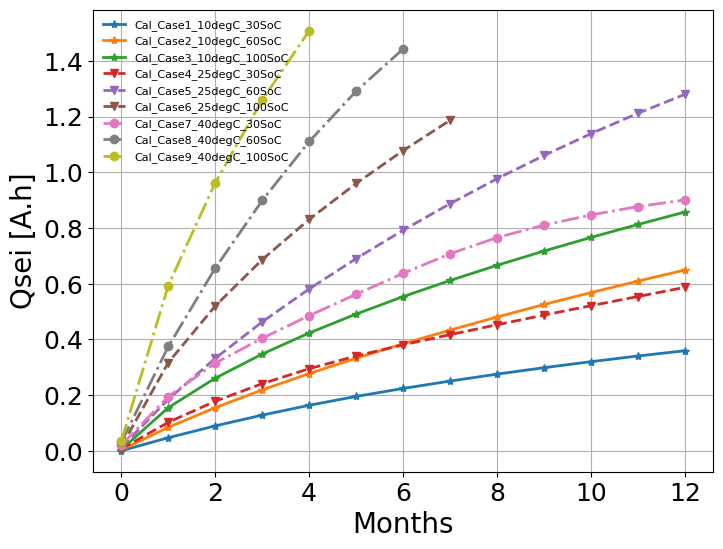}}
    \caption{a) Simulated calendar ageing trajectories of the cell at different  rest SoCs (30\%, 60\%, 100\%) and temperatures (10$^o$C, 25$^o$C, 40$^o$C).  Calendar ageing is higher at higher temperature and higher rest SoC. b)Qsei [A.h]: Loss of capacity due to SEI (Ah) for the calendar ageing cases.}

    \label{fig3_calednarAg}
\end{figure}

Calendar ageing is dependent on temperature, SoC/potential and duration.   Figure~\ref{fig3_calednarAg}a shows trajectories of simulated calendar ageing in the cell when subjected to different combinations of temperatures and rest SoCs for a period of 12 months. The simulations were carried out at 30\% SoC, 60\% SoC and 100\% SoC, and 10$^o$C, 25$^o$C and 40$^o$C. Among the simulated cases, the ageing is highest in the 40$^o$C-100\% SoC case and lowest in  10$^o$C-30\% SoC case, as expected, demonstrating the consistency of the model and its applicability. Calendar ageing is higher at higher temperature and higher rest SoC \cite{Broussely2001, dubarry2018calendar,keil2016calendar} due to dominance of SEI capacity loss. 

At 25$^o$C and 100\% rest SoC, the capacity dropped to below 80\% SoH in 5 months. At 40$^o$C and 100\% rest SoC, the capacity dropped to below 80\% SoH in just above 2 months. At 10$^o$C and 30\%  SoC, the capacity dropped to 93\% SoH in 12 months. At a given temperature, for example, 40$^o$C, the calendar ageing is higher at 100\% rest SoC and lower at 30\% rest SoC, confirming the dependence on rest SoC. 

Cal-Case6 at 25$^o$C and 100\%, Cal-Case5 at 25$^o$C and 60\%, and Cal-Case3 at 10$^o$C and 100\%  have higher capacity fade than Cal-Case7 at 40$^o$C and 30\%, indicating rest SoC has stronger influence on calendar ageing. The analysis suggests a combination of temperature and SoC has to be looked at for quantifying the loss.

The corresponding capacity loss due to SEI for all the calendar ageing cases simulated are presented in Figure~\ref{fig3_calednarAg}b.

\subsection{Combined cyclic and calendar ageing: temperature, DoD, C-rate, and SoC dependence}
The model is used to predict capacity fade under combined cyclic and calendar ageing with 81 combinations of temperature (10$^o$C, 25$^o$C, 40$^o$C), C-rate (0.1 C, 0.3 C and 1.0 C), rest SoC (10\%, 60\%, and 100\%) and DoD  (50\%, 70\%, and 90\%). The conditions chosen are arbitrary and thought to be relevant for the given high energy dense cell. The list of the combinations is provided in the Data Availability section.

The model is subjected to one cycle per day at a specified C-rate (charge/ discharge) for a given DoD, and left to be at rest at the specified SoC at the end of the cycle for the rest of the day. The cycling procedure, with longer rest periods, accounts for the influence of both cyclic ageing and calendar ageing. The procedure is thought to represent an use-case of an EV two-wheeler. The cell is bound to be operated between 4.2 to 2.5~V at all times. The ambient temperature is maintained at the specified value, with heat loss due to convection, with h = 10~W/(m2.K), as a boundary condition.  

The legend, for example, $Cyl\_Case 1\_10degC\_0.1C\_10SoC\_50DoD$ indicates cycling at 10$^o$C ambient temperature with 0.1C/0.1D current for 50\% DoD and rest at 10\% SoC for the rest of the day (one cycle per day). The profiles of a cycle for few cases are shown in the supplementary information.

\subsubsection{For combinations of the usage conditions}
\begin{figure}
    \centering
    \includegraphics[width=1\linewidth]{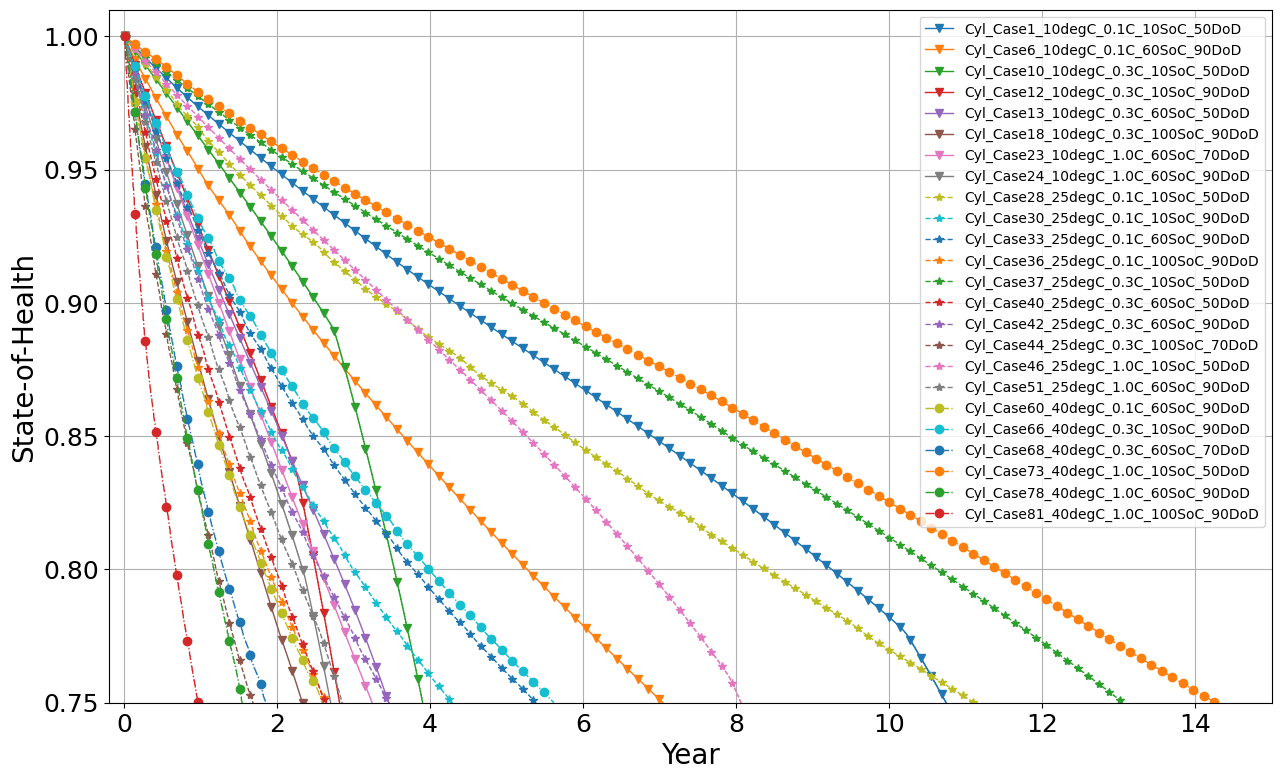}
    \caption{Simulated degradation trajectories of the cell for 24 cases (among the 81 cases simulated) of cycling at different combinations of temperature (10$^o$C, 25$^o$C, 40$^o$C), C-rate (0.1C, 0.3C and 1.0C), rest SoC (10\%, 60\%, and 100\%) and DoD  (50\%, 70\%, and 90\%). 1 cycle per day and at rest for the remaining period of the day at rest SoC. Cycle-life of the cell varies between 1 to 14 years to reach 75\% SoH under the specified usage conditions. }
    \label{fig1_26cases}
\end{figure}

Figure~\ref{fig1_26cases} shows the trajectories of capacity fade of the cell for 24 cases (arbitrarily chosen among the 81 cases simulated) with different combinations of temperature (10$^o$C, 25$^o$C, 40$^o$C), C-rate (0.1C, 0.3C and 1.0C), rest SoC (10\%, 60\%, and 100\%) and DoD  (50\%, 70\%, and 90\%), as indicated by the legends. The model predicts lifetime of the cell between 1 to 14 years to reach 75\% SoH, depending upon the specified usage conditions. 

The figure demonstrates the following observations of capacity fade in the cell. 
\begin{itemize}
    \item The trajectories can be sub-linear (Cyl\_Case 66), linear (Cyl\_Case 37) and super-linear (Cyl\_Case10) (accelerated, sudden death or knee type) based on the usage conditions. A cell can undergo any of these types of capacity fade, solely dependant on the usage conditions for a given chemistry. 
        
    \item For a few cases, the model predicts higher capacity fade:  at higher rest SoC, Cyl\_Case 36 > Cyl\_Case 33; for higher DoD, Cyl\_Case 30 > Cyl\_Case 28; for higher C-rate, Cyl\_Case 51 > Cyl\_Case 42 > Cyl\_Case 33; and for higher temperatures, Cyl\_Case 60 > Cyl\_Case 33 > Cyl\_Case 6. However, we may find contrary behaviour in a few other cases. For example, w.r.t temperature, Cyl\_Case 73 (40$^o$C) has longer lifetime than Cyl\_Case 46 (25$^o$C) while keeping rest of the  usage conditions the same.

    \item We cannot generalise the capacity fade trajectories, SoH and RUL to a single usage condition. Complex interlinking of coupled degradation mechanisms and their dominance, subjected to the combination of usage conditions and operating modes, has to be evaluated.
    
    \item Identical SoH for the same period of use with different usage histories doesn’t mean that the cell will have the same remaining useful life; at 4 years, Cyl\_Case 28 and Cyl\_Case 46 have the same SoH i.e. around 0.88, however, the remaining useful life after that instance is significantly different. We can make the observation in other cases too with cross-over in SoH. 
  
 \end{itemize}
 The closer version of the figure, up to 4 years, is presented in the Supplementary Information. 
 The simulated cycle-life data for the 81 cases are provided in the Data Availability section.   

\subsubsection{Competing effects between calendar and cyclic ageing}

\begin{figure}
    \centering
    \includegraphics[width=0.6\linewidth]{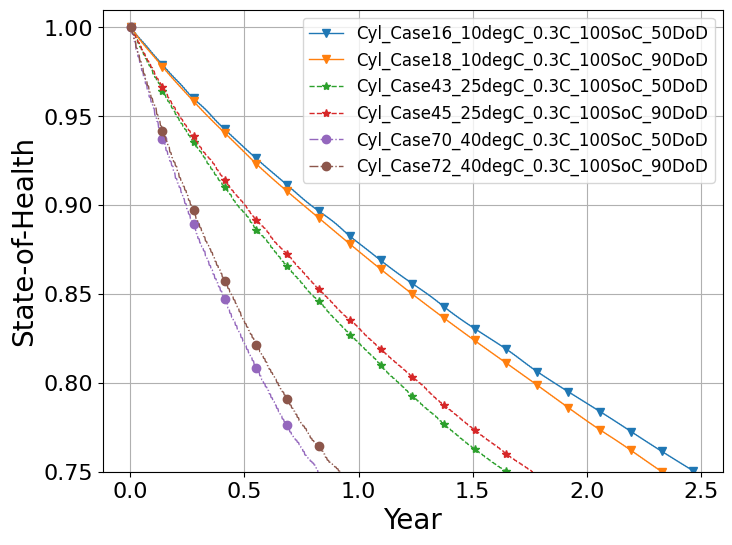}
    \caption{Influence of temperature (10$^o$C, 25$^o$C and 40$^o$C) and DoD (50\% DoD and 90\% DoD) on SoH during cycling with 0.3C and rest at 100\% SoC. The relationship between capacity fade and DoD isn't straightforward at different temperatures.  Capacity fade is faster for the 50\% DoD case compared to the 90\% DoD case, at 25$^o$C and 40$^o$C. At 10$^o$C, the case is different; capacity fade is faster in the 90\% DoD  case.}
    \label{fig2:compete1}
\end{figure}

\begin{figure}
    \centering
    \includegraphics[width=1\linewidth]{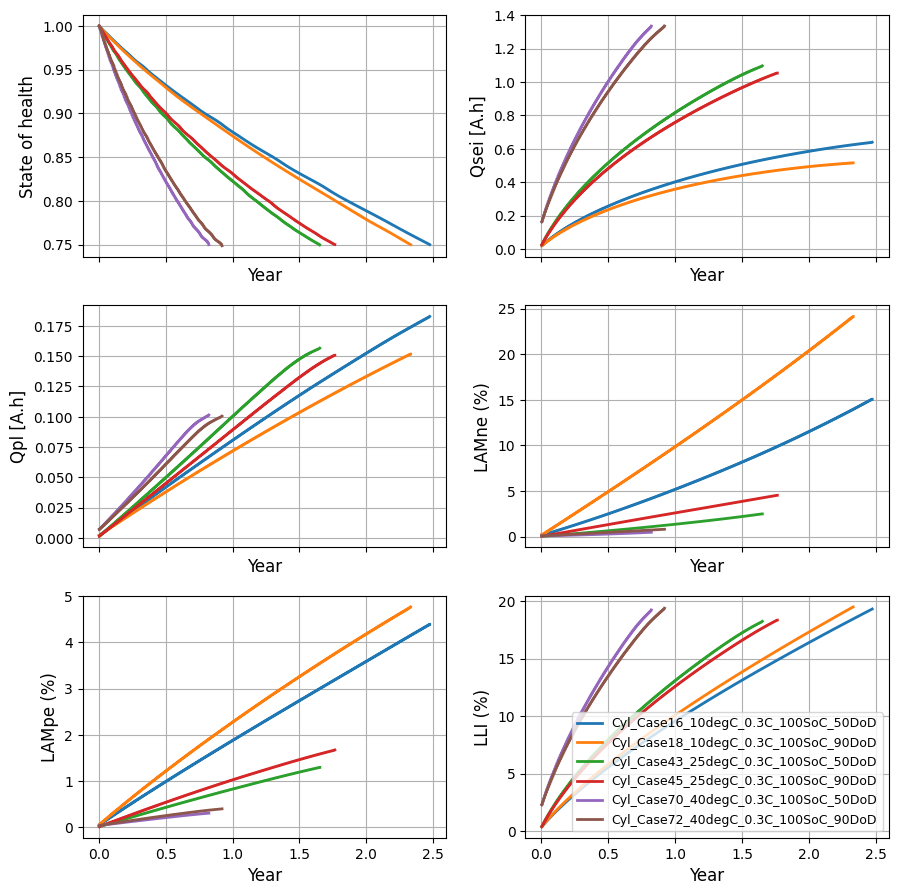}
    \caption{Simulated degradation mode analysis for the 6 cases considered in Figure~\ref{fig2:compete1}. Capacity fade is faster for the 50\% DoD case compared to the 90\% DoD case, at 25$^o$C and 40$^o$C. At 10$^o$C, the case is different; capacity fade is faster in the 90\% DoD  case. Qsei: Capacity loss due to SEI (Ah), Qpl: capacity loss due to Li plating (Ah), LAMne: LAM in negative electrode (\%), LAMpe: LAM in positive electrode (\%), and  LLI: loss of lithium inventory (\%).  }
    \label{fig3:competeDMA}
\end{figure}
Figure~\ref{fig2:compete1} shows the effect of temperature on cycle-life at 10$^o$C, 25$^o$C and 40$^o$C. The figure also shows the effect of DoD at each temperature: 50\% DoD and 90\% DoD. For all the six cases shown in the figure, the cell is cycled at  0.3C/0.3D and kept at 100\% SoC during rest for the day after the cycle. The cycle-life for the cases is between 0.7 to 2.5 years. 

The figure shows higher capacity fade at higher temperature due to dominance of the SEI capacity loss, as expected. However, on the other hand, 
% we expect higher capacity fade at higher DoD for a given temperature, but the case is different. 
the capacity fade is faster for the 50\% DoD case compared to the 90\% DoD case, at 25$^o$C and 40$^o$C. At 10$^o$C, the case is different; capacity fade is faster in the 90\% DoD case.  

The reasons for the behaviour as follows. Note that the cell is kept at rest 100\% SoC after the cycling. The cycling for 50\% DoD takes less time than the cycling for 90\% DoD. The difference in time period for the cycling between these cases at 0.3~C is been added to the rest period of the day, during which calendar ageing is in operation. At 25$^o$C and 40$^o$C, the calendar ageing during rest for the 50\% DoD cases is dominating the cyclic ageing of the 90\% DoD cases, leading to higher capacity loss. On the contrary, at 10$^o$C, the calendar ageing effects are minimal compared to the 25$^o$C and 40$^o$C cases, and the cycling ageing  due to higher LAM is dominant at 90\% DoD than the 50\% DoD case. So we see higher capacity fade at 90\% DoD.

The corresponding DMA simulations are shown in Figure~\ref{fig3:competeDMA}. The figure confirms the hypothesis; the capacity loss due to SEI ($Qsei$) is higher in the 50\%DoD cases than the 90\%DoD cases at 25$^o$C and 40$^o$C. And, the SEI loss (Ah) for both the temperatures is far higher than the loss at 10$^o$C, indicating the dominance of SEI w.r.t temperature. The total loss due to Li plating (Ah), though faster ($Qpl$) at higher temperatures, is far less than the SEI loss, confirming the dominance of SEI loss at higher temperatures. The higher rate for lithium plating is a manifestation due to the higher over-potential contribution by the growing SEI layer thickness.  

On the other hand, the figure shows LAMne (\%) and LAMpe (\%) are dominant at 10$^o$C and a few times higher than the cases at 25$^o$C and 40$^o$C. Loss due to LAMne is substantially higher for the 90\% DoD case than for the 50\% DoD case at 10°C, and markedly exceeds that of the all higher-temperature cases.  At low temperatures, the LAM losses are higher as stresses generated are higher with higher concentration gradients within the particles , due to slower Li diffusion. The LAMne is several times higher than the LAMpe though, as expected. The LLI loss (\%), from the SEI and plating combined, is also shown for all the cases.

\subsubsection{Dependence on rest SoC and C-rate}
\begin{figure}
    \centering
     
    \subfloat{}{\includegraphics[width=0.48\linewidth]{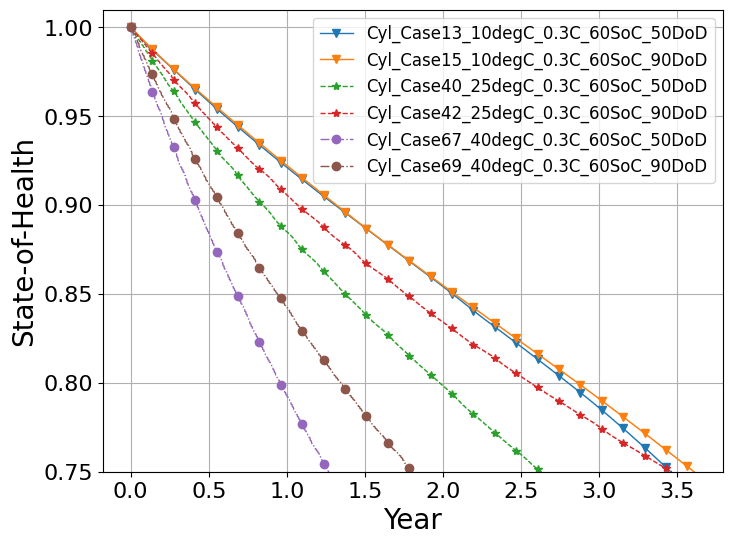}}
    \subfloat{}{\includegraphics[width=0.48\linewidth]{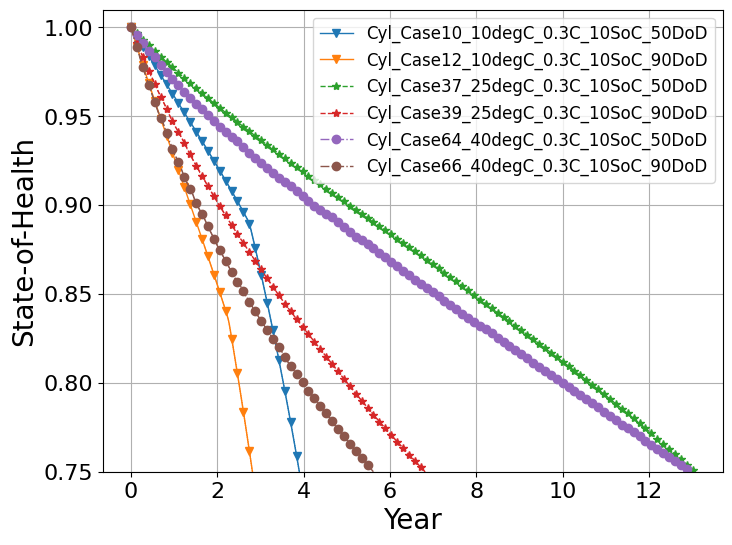}}
    \subfloat{}{\includegraphics[width=0.48\linewidth]{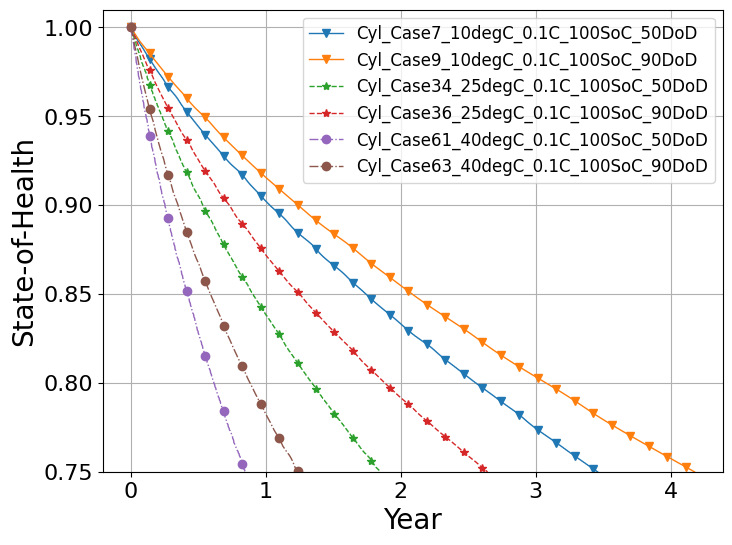}}
    \subfloat{}{\includegraphics[width=0.48\linewidth]{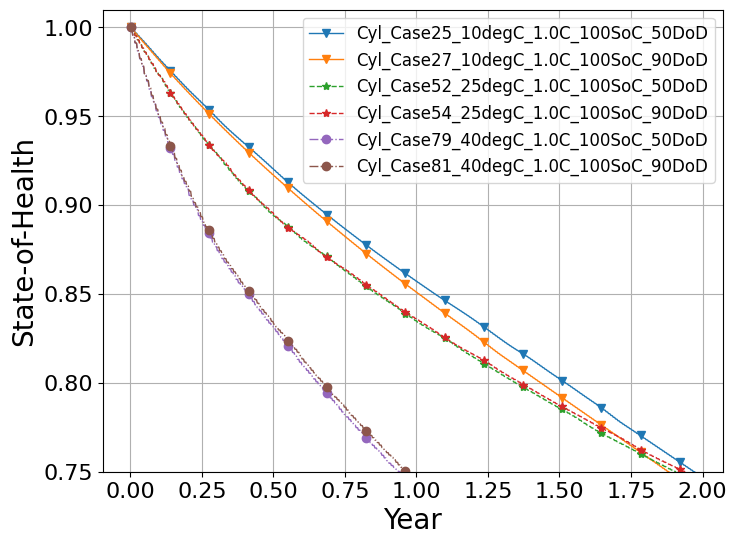}}
    \subfloat{}{\includegraphics[width=0.48\linewidth]{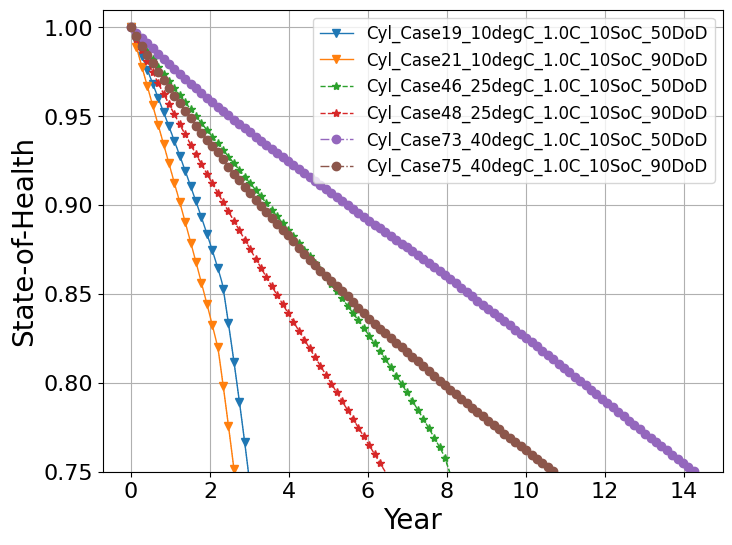}}
    \subfloat{}{\includegraphics[width=0.48\linewidth]{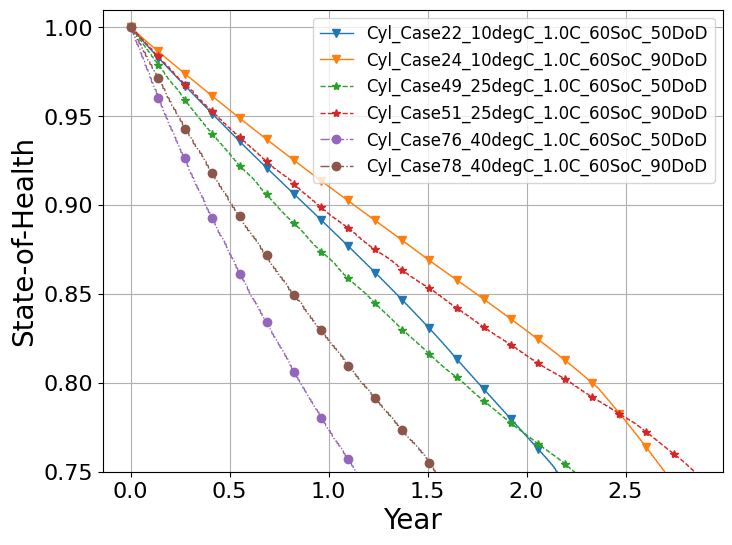}}
    
    \caption{Simulated ageing trajectories for the cell at different  rest SoCs (10\%, 60\%, 100\%) and C-rates (0.1, 0.3 and 1) at 10$^o$C, 25$^o$C, and 40$^o$C; and 50\% DoD and 90\% DoD . The trajectories are the extended studies to the cases presented in Fig.~\ref{fig2:compete1} and has to be seen in comparison with it. a \& b) the effect of change in rest SoC to 60\% and 10\% on the cycle-life while keeping other conditions the same. c \& d) the effect of change in C-rate to 0.1C and 1C while fixing the remaining conditions the same. e) the effect of change in C-rate to 1C and the SoC to 10\%. f) the effect of change in C-rate to 1C and SoC to 60\%  on the cycle-life. }

    \label{fig3_severalcases}
\end{figure}
  Figure~\ref{fig3_severalcases} shows the effect of rest SoC and C-rate during cycling for extended cases with respect to the cases discussed in Fig.~\ref{fig2:compete1}. While keeping the temperature and DoD the same, we varied rest SoC and C-rate to predict the effect on cycle-life and see the difference from the cases presented earlier. 

   Fig.~\ref{fig3_severalcases}a and b show the effect of change in rest SoC to 60\% and 10\% on the cycle-life while keep other conditions the same. Fig.~\ref{fig3_severalcases}c and d show the effect of change in C-rate to 0.1C and 1C. Fig.~\ref{fig3_severalcases}e shows the effect of change in C-rate to 1C and SoC to 10\%. Fig.~\ref{fig3_severalcases}f shows the effect of change in C-rate to 1C and SoC to 60\% on the cycle-life.  The cycle-life varies between 0.85 to 14 years. The simulated DMA for the each case is presented in the Supplementary Information.

   Fig.~\ref{fig3_severalcases}a shows, at the 60\% SoC, the calendar ageing effects in the 50\% DoD cases still dominate the cycling ageing effects in the 90\% DoD cases at higher temperatures. The cycle-life for all the cases is longer than the respective cases shown in Fig.~\ref{fig2:compete1} due to lower rest SoC, and is between 1.25 to 3.6 years. Contrary to the previous result, the cycle-life at 10$^o$C for both the cases are closer to the 90\% DoD case at 25$^o$C. 
   
   Fig.~\ref{fig3_severalcases}b shows that, at the 10\% SoC, the calendar ageing effects are far lower than the cycling effects and the cycle-life at the 50\% DoD for 25$^o$C and 40$^o$C has extended beyond 12 years. The 90\% DoD cases have shorter life than the 50\% DoD cases, at all the temperatures. Moreover, the cycle-life for all the cases are higher than the respective cases shown in Fig.~\ref{fig2:compete1} and varies between 2.6 to 13 years. Moderate increase in cycle-life at 10$^o$C and longer life at 25$^o$C and 40$^o$C can be observed, for lowering the rest SoC. 

   Fig.~\ref{fig3_severalcases}c shows, at 0.1C, similar results at 25$^o$C and 40$^o$C w.r.t to DoD as Fig.~\ref{fig2:compete1}, indicating calendar ageing effects still dominate. However, at 10$^o$C, contrary to the previous observation, the cyclic effects are not dominant and the 50\% DoD case has shorter life than the 90\% DoD, indicating the role of low C-rate. Cycle-life is moderately higher than the cases shown in the figure and is between 0.8 to 4.2 years with significant change at 10$^o$C.

   Fig.~\ref{fig3_severalcases}d shows, at 1C, almost equal cycle-life at 25$^o$C for both the DoDs, indicating a balance between the calendar and cycling effects.  At 40$^o$C, the cell has shortest life, due to the 100\% rest SoC contributions. The cycle-life for the cases varies between 0.9 to 2 years. 

   Fig.~\ref{fig3_severalcases}e shows, at 1C and 10\% SoC, that the 40$^o$C cases have longer cycle life between 10 to 14 years due to reduced calendar ageing effects at 10\% SoC, capturing the influence of rest SoC/voltage at high temperatures. The cycling effects of 1C at 40$^o$C  have little effect on cycle-life due to enhanced Li diffusion in the particles. The cases at 10$^o$C have shorter life than the cases at other temperatures, showing contrary observations than Fig.~\ref{fig2:compete1}, highlighting the role of rest SoC. The 90\% DoD cases have shorter life than the 50\% DoD cases at the three temperatures, indicating the role of C-rate at reduced SoC. The cycle-life for the cases varies between 1.5 to 14 years.

   Fig.~\ref{fig3_severalcases}f, at 1C and 60\% SoC, shows that the cases at 25$^o$C and 10$^o$C have similar cycle-life and longer than at 40$^o$C. The calendar ageing effects for longer rest period are still dominant at 1C for all the three temperatures, and the 50\%DoD cases have shorter life than the 90\%DoD cases. The 50\%DoD case at 40$^o$C has shorter cycle-life than the other cases. The cycle-life for the cases varies between 1.1 to 2.8 years. 

   % \uline{The corresponding degradation mode analysis simulation results for each case shown in Fig.~\ref{fig3_severalcases} are presented in the supplementary information.} 

In summary, the interplay between the coupled mechanics in response to the usage conditions for a given operating mode would decide the cycle-life and warrants consistent parametrisation of degradation mechanisms for prediction of cycle-life.  It is also to be understood that the model results validated against experimental data of a single set of usage conditions won't be sufficient for extension to other usage conditions.  Based on the insights from the current study, validation of the model's SoH predictions with experimental data will be pursued in future work. \citet{kirkaldy2024lithium} work is a good source of experimental data for LG M50 cell for a few usage conditions. 
 
The ability to predict consistent capacity fade trajectories for different usage conditions highlights the superiority of physics-based models over other class of models for predicting SoH and RUL. Simulated degradation mode analysis results (Qsei[A.h], Qpl [A.h], LAMne(\%), LAMpe(\%) and LLI (\%) along with the SoH) for the 81 cycling cases, 9 calendar ageing cases and 3 standard cycling cases, considered in this study, are provided in the Data Availability section for further analysis. The data can be used for training physics-informed data-driven models for suitable applications, as generating the experimental data for such vast cases isn't easy. 
 
 \section{Conclusion}

An accurate estimation of SoH and RUL of lithium-ion cells in the applications (EV, energy storage, consumer electronics) facilitates enhanced safety, improves user experience, and helps in preventive maintenance. The work attempts to bridge the gap between the modelling and experiments of lithium-ion batteries to develop an accurately parameterised model. It provides strategies to parameterise a model based on DMA data and vouches for its consistency check in different operating modes.
\begin{itemize}
\item We built an electrochemical-thermal-degradation model for the LG M50 lithium-ion cell using PyBaMM, with consistant parameterisation of SEI and lithium plating at the negative electrode and active material loss in both electrodes, based on the insights form the experimental DMA data for the cell presented in the literature\cite{kirkaldy2024lithium}. The parameters are tuned for the SEI domination at high temperatures, and LAM and Li plating domination at low temperatures. The mechanisms were made to be dependent on temperature, storage SoC, DoD and C-rate.

\item We predicted qualitative degradation trajectories under calendar ageing mode for 9 combinations of temperature and rest SoC,  for an year. Predicted the loss of capacity due to SEI for the rest conditions at temperature: 10$^o$C, 25$^o$C, and 40$^o$C and rest SoC: 10\%, 60\%, and 100\%. The predictions demonstrated that the capacity loss is higher at higher temperature and higher SoC.

\item We simulated qualitative degradation trajectories for 81 combinations of rest SoC, DoD, C-rate and temperature, under combined cyclic and calendar ageing modes. The degradation mode simulations quantify the contributions from SEI, Li plating and LAM in negative and positive electrodes to the total loss. The variations considered in usage conditions are, for rest SoC: 10\%, 60\%, and 100\%; for C-rate: 0.1C, 0.3C, and 1C; for DoD: 50\%, 70\%, and 90\%; and for temperature: 10$^o$C, 25$^o$C, and 40$^o$C. 

\item The model predicted degradation trajectories which are sub-linear, when the cell is operated gently; linear under moderate conditions; and super-linear (accelerated capacity fade/knee-type) , when the cell is operated aggressively, based  on the usage conditions, for the given chemistry. The cell cycle-life varies between 0.85 to 14 years, for 75\% SoH.  

\item We cannot generalise the capacity fade trajectories, SoH and RUL to a particular usage condition, during combined cyclic and calendar ageing. The competing effects between calendar and cycling ageing are demonstrated for different usage conditions. The predictions highlight the complex interlinking of coupled degradation mechanisms and clearly indicate their dominance regime, subjected to the combination of usage conditions and operating modes.

\item The work quantified the link between degradation mechanisms, operating conditions and predicted cycle-life. 

\item The simulated SoH and degradation mode analysis data generated under the 81 cyclic ageing use cases, the 9 calendar ageing cases and 3 standard cycling conditions are made available online for training physics-informed data driven models. 

\item Validation of the model's degradation trajectories with experimental data with accurate parameterisation will be pursued in future work. 

% \item Demonstration of consistency in calendar ageing and cyclic ageing together using a DFN model is presented for the first time, as per our knowledge. 

\end{itemize}

% \section*{Nomenclature}
%% The Appendices part is started with the command \appendix;
%% appendix sections are then done as normal sections

\section*{Supplementary information}
\label{sec:supple}
The supplementary information to this article can be found online at: \href{https://zenodo.org/records/19491921?preview=1&token=eyJhbGciOiJIUzUxMiJ9.eyJpZCI6IjdmMDkyZTJhLTQ4OWEtNDliNS04MmU3LTMzMzVmNjc2MWVmZCIsImRhdGEiOnt9LCJyYW5kb20iOiI3YzMzYWRkOTE4NWRiYTM4MmZjNWYxMGY3YTg4N2E5MCJ9.1zIOSvI2K5sDyaqivNg2napPYvvFg0xqwvg7O8SedPZhpumO1xiKdGDAqVLkprfyUMpWdJyyIuqK45TCrMb94Q}{supplementaryInfo-and-data-linking-Zenodo-GM-Apr2026}.

%\href{https://tinyurl.com/supplementaryI-GM-Apr2026}{https://tinyurl.com/supplementaryI-GM-Apr2026}..

\section*{Funding}
The author acknowledges the funding from IIT (BHU) Seed Grant for enabling this work. 

\section*{Data availability}
The lists of usage cases, and the simulated data of SoH and internal degradation modes for the 81 usage cases, the 9 calendar ageing cases and the 3 standard cycling cases  are available for download here: \href{https://zenodo.org/records/19491921?preview=1&token=eyJhbGciOiJIUzUxMiJ9.eyJpZCI6IjdmMDkyZTJhLTQ4OWEtNDliNS04MmU3LTMzMzVmNjc2MWVmZCIsImRhdGEiOnt9LCJyYW5kb20iOiI3YzMzYWRkOTE4NWRiYTM4MmZjNWYxMGY3YTg4N2E5MCJ9.1zIOSvI2K5sDyaqivNg2napPYvvFg0xqwvg7O8SedPZhpumO1xiKdGDAqVLkprfyUMpWdJyyIuqK45TCrMb94Q}{supplementaryInfo-and-data-linking-Zenodo-GM-Apr2026}.

%\href{https://tinyurl.com/data-linking-GM042026}{https://tinyurl.com/data-linking-GM042026}. 

\section*{CRediT statement}
G. Madabattula: Conceptualisation, methodology, formal analysis, investigation, visualisation, writing-original draft preparation and review and editing, funding, project administration. 

\section*{ORCID ID}
Ganesh Madabattula: https://orcid.org/0000-0001-7915-0770  \\

\appendix

\section{}
\subsection{Beginning-of-life C-rate performance of the cell}
\label{sec:append1}

\begin{figure}[htbp]
\centering
\subfloat{}{\includegraphics[width=0.48\linewidth]{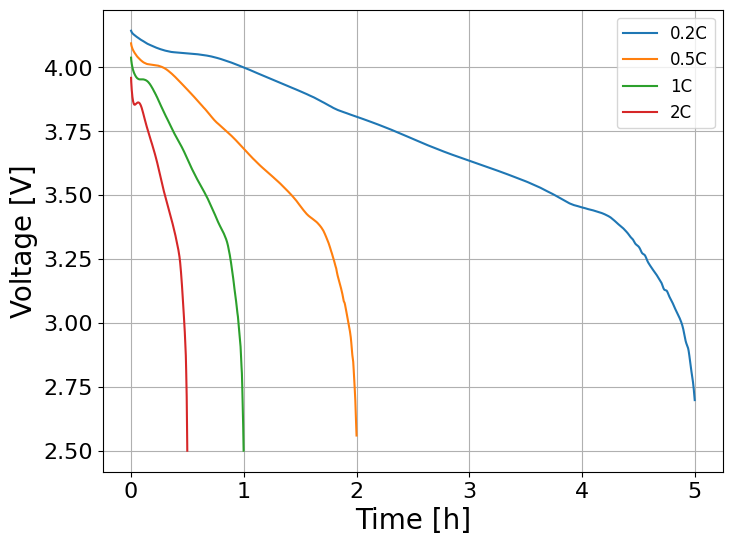}}
\subfloat{}{\includegraphics[width=0.48\linewidth]{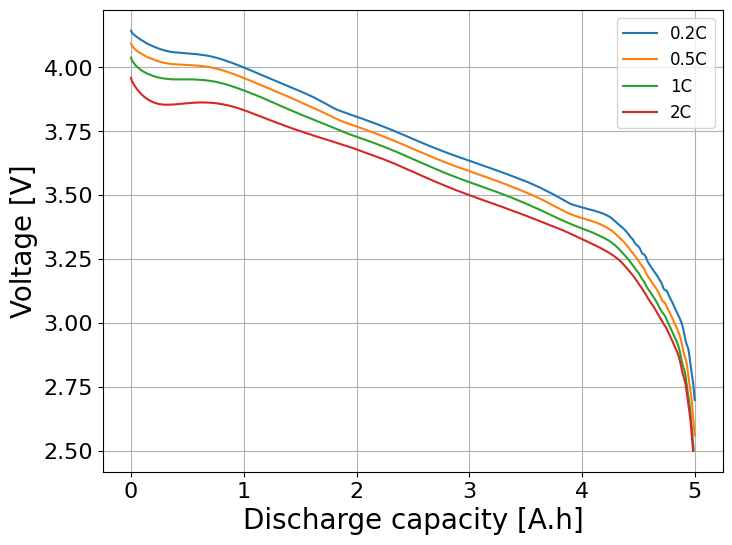}}
\caption{Simulations of beginning-of-life behaviour of the cell at different C-rates. Voltage vs. time  and voltage vs. discharge capacity, at 0.2C, 0.5C, 1C and 2C.}
\label{fig1apxBOL}
\end{figure}

Figure~\ref{fig1apxBOL} shows beginning-of-life behaviour of the cell at different C-rates. The figure shows consistent profiles of voltage vs. time  and voltage vs discharge capacity, at different C-rate and confirms the cell capacity.

%% If you have bib database file and want bibtex to generate the
%% bibitems, please use
%%
%%  \bibliographystyle{elsarticle-num-names} 
%%  \bibliography{<your bibdatabase>}

%% else use the following coding to input the bibitems directly in the
%% TeX file.

%% Refer following link for more details about bibliography and citations.
%% https://en.wikibooks.org/wiki/LaTeX/Bibliography_Management

\bibliography{references_knee}
%\bibliography{doc/references}
% \begin{thebibliography}{00}

% %% For authoryear reference style
% %% \bibitem[Author(year)]{label}
% %% Text of bibliographic item

% \bibitem[Lamport(1994)]{lamport94}
%   Leslie Lamport,
%   \textit{\LaTeX: a document preparation system},
%   Addison Wesley, Massachusetts,
%   2nd edition,
%   1994.

% \end{thebibliography}
\end{document}